\documentclass[aps,prl,superscriptaddress,twocolumn,longbibliography,floatfix]{revtex4-2}

\usepackage{amsfonts}
\usepackage{amsmath}
\usepackage{mathrsfs}
\usepackage{amssymb}
\usepackage{graphicx}
\usepackage{float}
\usepackage{comment}
\usepackage{bm}
\usepackage{color}
\usepackage[colorlinks=true,linkcolor=blue,anchorcolor=blue,citecolor=blue,urlcolor=blue]{hyperref}
\usepackage[normalem]{ulem}
\usepackage{siunitx}
\usepackage[tracking]{microtype}
\usepackage[version=4]{mhchem}
\usepackage{elements}
\usepackage{soul}
\setstcolor{red}

\SetTracking{ encoding = *, shape = sc }{ 25 }

\newcommand{\ket}[1]{\lvert #1 \rangle}

\DeclareMathSymbol{\shortminus}{\mathbin}{AMSa}{"39}

\newcommand{\RomanNumeralCaps}[1]{\MakeUppercase{\romannumeral #1}}

\begin{document}

\title{Macroscopic Self-Trapping and Dynamical Phase Transition in Momentum Space Bose-Einstein Condensates}
\author{Colby Schimelfenig}
\affiliation{Department of Physics and Astronomy, Washington State University, Pullman, WA 99164-2814, USA}
\author{Federico Serrano}
\affiliation{Department of Physics and Astronomy, Washington State University, Pullman, WA 99164-2814, USA}
\author{Corey Halverson}
\affiliation{Department of Physics and Astronomy, Washington State University, Pullman, WA 99164-2814, USA}
\author{Annesh Mukhopadhyay}
\thanks{}
\affiliation{Department of Physics and Astronomy, Washington State University, Pullman, WA 99164-2814, USA}
\author{Qingze Guan}
\email{qingze.guan@wsu.edu}
\affiliation{Department of Physics and Astronomy, Washington State University, Pullman, WA 99164-2814, USA}
\author{Peter Engels}
\email{engels@wsu.edu}
\affiliation{Department of Physics and Astronomy, Washington State University, Pullman, WA 99164-2814, USA}

\begin{abstract}
Self-trapping is a hallmark phenomenon of nonlinear dynamics.
It has significant applications in modern physics, including band structure engineering, phase transition dynamics, quantum metrology, and more. Dilute-gas Bose-Einstein condensates (BECs), in which self-trapping can arise from interatomic interactions, are a prime testbed for probing nonlinear dynamics. In this Letter, we report the observation of self-trapping in a spin-orbit coupled BEC subjected to a stationary optical lattice. We employ Raman-induced spin-orbit coupling, complemented by a matching optical lattice that facilitates coupling between momentum eigenstates of the spin-orbit coupled system. By ramping the Raman detuning, we probe atomic current flow between these eigenstates and identify a clear distinction between a delocalized mixed state and a self-trapped regime. 
Following a quench of the Raman detuning, the time-averaged atomic current exhibits non-analytic behavior across the transition between these two regimes in certain parameter ranges, signaling a dynamical phase transition in the system.
\end{abstract}

\maketitle

\paragraph{\textcolor{blue}{Introduction.}}
Macroscopic quantum self-trapping (MQST) is a nonlinear phenomenon intrinsic to Bose-Einstein condensates (BECs), emerging when interatomic interactions dominate the dynamics~\cite{smerzi1997, raghavan1999, milburn1997}. It manifests as a self maintaining population imbalance between quantum states,  where tunneling would otherwise occur in the absence of a nonlinearity. MQST has been studied theoretically across a variety of systems, including double wells~\cite{cui2010, fu2006, ottaviani2010, pudlik2014, ruostekoski1998, wang2006a, wang2018, xiong2009, julia-diaz2010, khabdullaev2023}, optical lattices~\cite{wang2006, xue2008}, and driven systems~\cite{kolovsky2010, wuster2012}. Experimental observations have also been reported in exciton polaritons~\cite{abbarchi2013} and real-space BEC dynamics~\cite{albiez2005, anker2005, levy2007}.

As a fundamental feature of many-body self-interacting systems, MQST plays a central role in applications such as Josephson junctions~\cite{josephson1962, josephson1974, abbarchi2013, albiez2005, julia-diaz2010, khabdullaev2023, levy2007, raghavan1999, xiong2009},  and is closely linked to dynamical phase transitions (DPTs)~\cite{fu2006, julia-diaz2010, muniz2020}. DPTs describe qualitative changes in the long-term dynamics of a system as input parameters are varied~\cite{marino2022, corps2023}. DPTs have been studied in a variety of systems including, for instance, qubit quantum simulators~\cite{zhang2017}, open quantum systems~\cite{diehl2010, klinder2015}, driven dissipative systems~\cite{espigares2013}, and beyond~\cite{ramos2007, corps2023}. Their relevance extends even further, with applications in fields such as complex networks ~\cite{ni2019} and cancer research~\cite{davies2011}. The type of DPT focused on in this work, also called a classical DPT or DPT-\RomanNumeralCaps{1}~\cite{corps2023}, falls under the broad umbrella of non-equilibrium phase transitions and is different from dynamical quantum phase transitions~\cite{heyl2018, roggero2021, vajna2015, abdi2019, canovi2014, heyl2013, karrasch2013, vajna2014, heyl2017}. All references to DPTs below are meant to refer to a DPT-\RomanNumeralCaps{1}.

Motivated by previous studies of MQST and DPT in real-space BECs, this Letter establishes a new highly tunable experimental platform by exploiting a momentum-state coupled BEC. To achieve this, we utilize a Raman-induced spin-orbit coupled (SOC) BEC, in which atomic linear momentum is coupled to internal spin states via Raman transitions~\cite{ho2011, jimenez-garcia2015, lin2011, zhang2016}. Complementing the two Raman laser beams is a stationary optical lattice that enhances coupling between the two minima of the SOC ground-state band structure. The effects of optical lattices applied to BECs is well established~\cite{morsch2006, guan2020, shaw2023}. In our setup, we create a matching optical lattice (ML) by choosing the experimental parameters such that the lattice connects the two momentum states located at the minima of the double-well dispersion of the lowest SOC band. This configuration facilitates the creation of two-state spin mixtures by inducing tunneling between momentum eigenstates. 

The intensity of the ML provides an experimental handle to tune the dynamical regime of the system. Increasing the ML intensity promotes delocalized mixed-state dynamics, while decreasing it drives the system towards the MQST regime. With this paradigm, we present experimental evidence of MQST by mapping state transfer dynamics under a linear detuning ramp of the Raman beams. Building on this, we further extend our investigation to include the observation of a DPT in the SOC + ML system. We probe the DPT by quenching the Raman detuning, which imparts energy to the system and induces oscillatory behavior in momentum-state populations. Crucially, the long-term dynamics differ significantly depending on whether the system resides in the MQST regime or the delocalized mixed state, consistent with the defining features of a DPT.

The experimental observations are well described by two theoretical approaches. We first employ a two-mode Josephson Junction model, wherein the condensate is treated as a superposition of two dressed states localized in a momentum-space double well. To incorporate finite-size effects, we additionally perform Gross-Pitaevskii (GP) simulations. Both approaches accurately capture the characteristic features of the MQST and the critical behavior of the DPT observed experimentally. However, in certain parameter regimes, spatial excitations arising from finite-size effects modify the critical behavior of the DPTs, revealing dynamics beyond the two-mode model. Investigating the coupling between the spin and spatial degrees of freedom in the system presents an intriguing direction for future research.

\begin{figure}[t]
\centering
\includegraphics[scale=1.0]{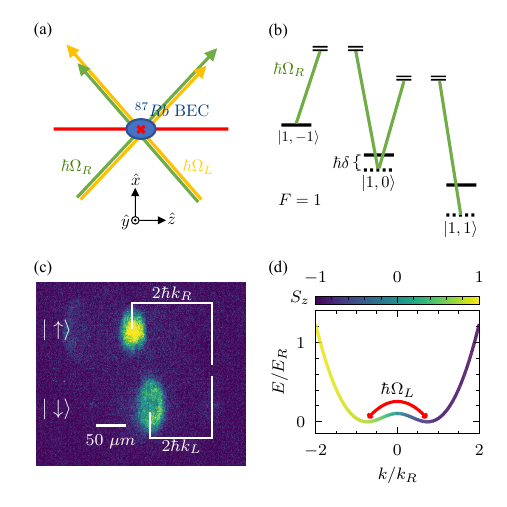}
\caption{Experimental setup for the SOC + ML system. (a) Diagram of laser configuration. BEC (blue oval) is held by a cross-dipole trap (red line and x). Raman and optical lattice beams are shown by the green and yellow beams intersecting with the BEC. (b) Schematic state diagram for Raman coupling between different $m_{F}$ states within the F=1 hyperfine manifold of the $^{87}$Rb $5^{2}S_{1/2}$ ground state. (c) Example image of SOC + ML BEC taken using time of flight and Stern-Gerlach imaging. (d) The lowest band in the single-particle dispersion relation of the SOC system for $\hbar \Omega_{R}=\qty{2.7}{E_{R}}$ and $\delta=2\pi\times\qty{0}{Hz}$. Coupling between the two band minima is enhanced by a weak optical lattice with Rabi coupling $\hbar\Omega_L$.
}
\label{fig:ExpSetup}
\end{figure}

\paragraph{\textcolor{blue}{Theoretical background.}}
To model the SOC coupling, we consider the three hyperfine states within the $F=1$ manifold of $^{87}$Rb, as shown schematically in Fig.~\ref{fig:ExpSetup}(b). In the presence of a magnetic field, the Raman lasers near resonantly couple the $|1,-1\rangle$ and $|1,0\rangle$ states which we label as $\ket{\uparrow}$ and $\ket{\downarrow}$, respectively. The $|1,1\rangle$ state, which is far detuned due to the quadratic Zeeman shift $2\hbar q_z$, can be adiabatically eliminated, leading to an effective shifted detuning $\hbar\delta$ for the resulting spin-$1/2$ model~\cite{sup_mat}.  

Within the mean-field approximation, the dynamics of the spin-$1/2$ system are described by the GP equation
\begin{widetext}
\begin{multline}
\label{eq:gp_equation}
    i\hbar\frac{\partial\Psi(\mathbf{r}, t)}{\partial t} 
    = \Bigg[H_{\text{SOC}} + V_L(z)\sigma_0 + U(\mathbf{r})\sigma_0
    +
    \begin{pmatrix}
    g_{\uparrow\uparrow}|\psi_{\uparrow}(\mathbf{r},t)|^2+g_{\uparrow\downarrow}|\psi_{\downarrow}(\mathbf{r},t)|^2 & 0\\
    0 & 
    g_{\uparrow\downarrow}|\psi_{\downarrow}(\mathbf{r},t)|^2+g_{\downarrow\downarrow}|\psi_{\downarrow}(\mathbf{r},t)|^2
    \end{pmatrix}\Bigg] \Psi(\mathbf{r}, t),
\end{multline}
\end{widetext}
where the wave function $\Psi(\mathbf{r}, t)=[\psi_\uparrow(\mathbf{r}, t), \psi_\downarrow(\mathbf{r}, t)]^T$ is normalized to the total number of atoms, $\int d^3\mathbf{r}|\Psi({\bf{r}},t)|^2=N$, $\sigma_0$ is the 2-by-2 identity matrix, $U(\mathbf{r})$ is the harmonic trap potential, $V_L(z)=\hbar\Omega_L\sin^2(k_Lz)$ is the optical lattice potential that couples states of the same spin that differ in momentum by $2l\hbar k_L\ (l=\pm 1, \pm 2,\dots)$ in the $z$-direction, and $g_{ss'}=4\pi\hbar^2a_{ss'}/m$ is the nonlinear coupling constant between spin-$s$ and spin-$s'$ states, with $m$ and $a_{ss'}$ denoting the atomic mass and the corresponding $s$-wave scattering lengths, respectively. The single-particle SOC Hamiltonian is given by
\begin{align}
    H_\text{SOC} = \frac{(-i\hbar \nabla \sigma_0 + \hbar k_R\mathbf{e}_z\sigma_z)^2}{2m} - \frac{\hbar\delta\sigma_z}{2}+\frac{\hbar \Omega_R\sigma_x}{2},
    \label{eq:soc_hamiltonian}
\end{align}
where $\sigma_i$ are Pauli matrices in the $i$-direction, $\mathbf{e}_z$ is the unit vector in the $z$-direction, and the SOC parameters $k_R$, $\hbar\delta$, and $\hbar\Omega_R$ denote the SOC strength, detuning, and Raman coupling strength, respectively. Throughout this work, we use the recoil energy $E_R = \hbar^2 k_R^2 / (2m)$ as the unit of energy. 

The dispersion of the SOC Hamiltonian for $\hbar\Omega_R < 4E_R$ and vanishing $\delta$ is shown in Fig.~\ref{fig:ExpSetup}(d). It exhibits a double-well structure, with the states near the left and right minima corresponding to dressed spin states $\chi_l$ and $\chi_r$, respectively. Focusing on the ML case, we take $k_L = k_{R} [1-(\hbar\Omega_{R}/4E_R)^2]^{1/2}$, which matches the momentum difference between the two minima.

As derived in ~\cite{mukhopadhyay2024}, ignoring the spin-dependence in the $g_{ss'}$, introducing the relative (right-left) population imbalance $s_z = (N_r - N_l)/(N_r + N_l)$ and the relative phase $\theta = \theta_l - \theta_r$ where $N_{r/l}$ and $\theta_{r/l}$ denote the population and the phase of particles occupying the right/left state, respectively, the system reduces to an effective two-mode Josephson junction model,
\begin{eqnarray}
    H_\text{eff} = -\tfrac{1}{2}gn\chi^2s_z^2 - \hbar\Omega_L\chi\sqrt{1-s_z^2}\cos\theta - \hbar\delta s_z,
\end{eqnarray}
where $n$ is the mean particle density, $\chi=|\chi_l^{\dagger}\chi_r|$ is proportional to the inter-mode hopping amplitude between the two band minima, and $g$ is the spin-independent coupling constant~\cite{sup_mat}. 

\paragraph{\textcolor{blue}{Experimental setup.}}
A detailed description of our SOC + ML system is given in~\cite{bersano2019, mukhopadhyay2024}. Briefly, we begin by producing a $^{87}$Rb BEC in a crossed optical dipole trap. We then adiabatically turn on two 789~nm Raman beams. The beams intersect at the position of the BEC with an angle $\beta \approx \ang{90}$ relative to each other [Fig.~\ref{fig:ExpSetup}(a)]. They induce spin-orbit coupling characterized by a momentum transfer of $2\hbar k_R\sin{\left(\beta/2\right)}$ to the atoms. An external magnetic field of 10~Gauss is applied to lift the degeneracy of the Zeeman sublevels within the $F=1$ hyperfine manifold. This results in a linear Zeeman shift between states $|1,-1\rangle$ and $|1,0\rangle$ of approximately $h\times\qty{7}{MHz}$, which is matched by the detuning between the Raman beams except for a small, adjustable detuning $\delta$ [Fig.~\ref{fig:ExpSetup}(b)]. The quadratic Zeeman shift between the $|1,0\rangle$ and $|1,1\rangle$ states is $2\hbar q_z\sim h\times $14~kHz, making the transition between $|1,0\rangle$ and $|1,1\rangle$ far off-resonant. Thus the $|1,1\rangle$ hyperfine state does not get noticeably populated. To induce tunneling between spin-orbit states, a pair of \qty{1064}{nm} laser beams co-linear with the SOC Raman beams is added to the system [Fig.~\ref{fig:ExpSetup}(a)]. For the Raman strength of $\hbar\Omega_{R} = $ 2.7 $E_R$ used in the experiments, this lattice connects the momentum states located at the two dispersion minima of the lower spin-orbit band [Fig.~\ref{fig:ExpSetup}(d)].

To measure observables such as the hyperfine spin state and the momentum of the atoms, we employ Stern-Gerlach separation combined with time-of-flight imaging [Fig.~\ref{fig:ExpSetup}(c)]. From the resulting images, we count the number of atoms in the spin-up and spin-down states, denoted as $N_{\uparrow}$ and $N_{\downarrow}$, and calculate the spin-polarization $S_{z} = (N_{\uparrow}-N_{\downarrow})/(N_{\uparrow}+N_{\downarrow})$ which serves as the order parameter in the equilibrium phase diagram. In the homogeneous limit, the spin-polarization and the left–right population imbalance are found to be linearly related, $S_z = (k_L/k_R) s_z$~\cite{sup_mat}.

\begin{figure}[t]
\centering
\includegraphics[scale=1.0]{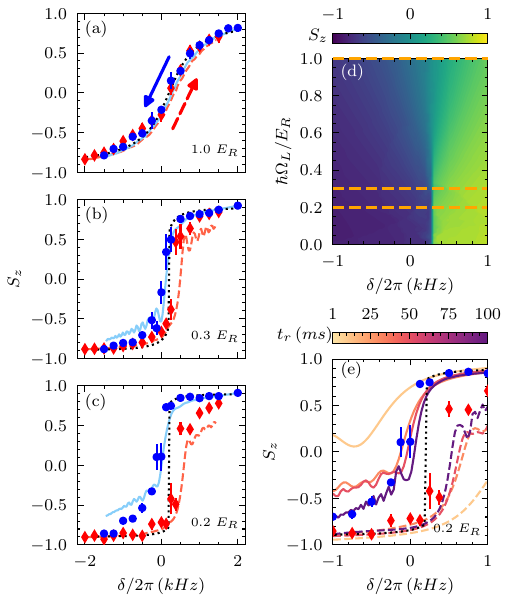}
\caption{Self-trapping dynamics. (a)-(c) Self-trapping spin dynamics on the ground state of the system following an adiabatic Raman detuning ramp. Panels (a), (b), and (c) correspond to $\hbar\Omega_{L}=$ 1.0, 0.3, and $\qty{0.2}{E_{R}}$ respectively. The blue dots (solid curves) show spin polarization of experimentally (numerically) acquired results at $\delta$ given a linear Raman detuning ramp from $\delta_i=2\pi\times\qty{5}{kHz}\rightarrow\delta_f=-2\pi\times\qty{1.5}{kHz}$ in $t_r = \qty{50}{ms}$. The red diamonds (dashed lines) signify experimental (numerical) results given a ramp from $\delta_i=-2\pi\times\qty{5}{kHz}\rightarrow\delta_f=2\pi\times\qty{1.5}{kHz}$. The blue solid (red dashed) arrow in panel (a) indicates the direction of the detuning ramp in the negative (positive) $\delta$ direction for all panels (a)-(c). (d) Ground state phase diagram of the SOC + ML system. Orange dashed lines show cuts across the spin-polarization phase transition at the three values of $\hbar\Omega_{L}$ seen in (a)-(c). (e) Same experimental data as in panel (c), for $t_r=\qty{50}{ms}$ and $\hbar\Omega_{L}=\qty{0.2}{E_{R}}$. Numerical solutions for different ramp times at $\hbar\Omega_{L}=\qty{0.2}{E_{R}}$ for $t_{r}$ ranging from $\qty{10}{ms}$ to $\qty{100}{ms}$ showing a convergence of dynamics at longer ramp times. Blue dots (red diamonds) are experimental data while solid curves (dashed curves) are GP simulations for negative (positive) detuning ramps. The black dotted lines in (a)-(c) and (e) represent solutions to the two-mode model.
}
\label{fig:GndStRamp}
\end{figure}

\paragraph{\textcolor{blue}{Self-trapping regime dynamics near phase transition.}}
Our first main results is the observation of self-trapping in a SOC + ML BEC (Fig.~\ref{fig:GndStRamp}). The ground state phase diagram shown in Fig.~\ref{fig:GndStRamp}(d), obtained via imaginary-time evolution of Eq.~\eqref{eq:gp_equation} and evaluation of $S_z$ at $gn=0.60\ E_R$ and $\chi=0.65$, shows a shift of the transition point from $\delta=2\pi\times\qty{0}{Hz}$ to a slightly larger value of $\delta\approx2\pi\times\qty{280}{Hz}$ at weak lattice strength, arising from the effective detuning induced by the $\ket{1,1}$ state. At equilibrium ($\dot{s_z}=\dot{\theta}=0$), the equations of motion $\dot{s_z}=-\partial_\theta H_\text{eff}/\hbar$ and $\dot{\theta}=\partial_{s_z} H_\text{eff}/\hbar$ define two regimes: for $\hbar\Omega_L < gn\chi$, the population imbalance switches discontinuously from positive to negative near $\delta=2\pi\times\qty{0}{Hz}$, while for $\hbar\Omega_L>gn\chi$ the transition is continuous. Experimentally, we probe this phase diagram by performing linear detuning ramps $\delta(t) = \delta_i + (\delta_f - \delta_i)t/t_r$ for fixed values of $\hbar\Omega_{L}$ along the dashed lines in Fig.~\ref{fig:GndStRamp}(d). In different experimental runs, the ramps are performed in two directions, going from a far detuned $\delta_{i}=2\pi\times\qty{5}{kHz} \rightarrow \delta_{f}=-2\pi\times\qty{1.5}{kHz}$ or $\delta_{i}=-2\pi\times \qty{5}{kHz} \rightarrow \delta_{f}=2\pi\times\qty{1.5}{kHz}$ in a ramp time $t_r = \qty{50}{ms}$.  Fig.~\ref{fig:GndStRamp}(a-c) show the spin-polarization dynamics for fixed values $\hbar\Omega_{L}$ of 1.0, 0.3, and $\qty{0.2}{E_{R}}$, respectively.

We observe a distinct shift in the spin dynamics among the three ramps, which originates from the system transitioning between two regimes: the self-trapping regime at low optical lattice strength and the delocalized (non-self-trapping) mixed-state regime at high optical lattice strength. For $\hbar\Omega_{L}=\qty{1.0}{E_{R}}$ shown in Fig.~\ref{fig:GndStRamp}(a), the spin polarization adiabatically follows the detuning ramp and does not depend on the ramp direction. In contrast, ramps with $\hbar\Omega_{L}=\qty{0.3}{E_{R}}$ [Fig.~\ref{fig:GndStRamp}(b)] and $\hbar\Omega_{L}=\qty{0.2}{E_{R}}$ [Fig.~\ref{fig:GndStRamp}(c)] show spin dynamics that are strongly influenced by the direction of the detuning ramp. For ramps with low optical lattice strength, the spin polarization exhibits hysteresis due to self-trapping, which can only be overcome with sufficiently large detuning. This effect becomes more pronounced at lower $\hbar\Omega_{L}$ as shown in the differences between Fig.~\ref{fig:GndStRamp}(b) and Fig.~\ref{fig:GndStRamp}(c). 

To ensure that the $\qty{50}{ms}$ long linear detuning ramps are quasi-adiabatic, such that the ramp rate does not affect the spin dynamics of the system, Fig.~\ref{fig:GndStRamp}(e) shows numerically determined curves representing the detuning-dependent spin polarization for different ramp times. For ramp times $t_{r} \geq \qty{50}{ms}$, the dynamics of the system are convergent, exhibiting negligible changes with further increases in ramp time. 

\begin{figure}[t]
\centering
\includegraphics[scale=1]{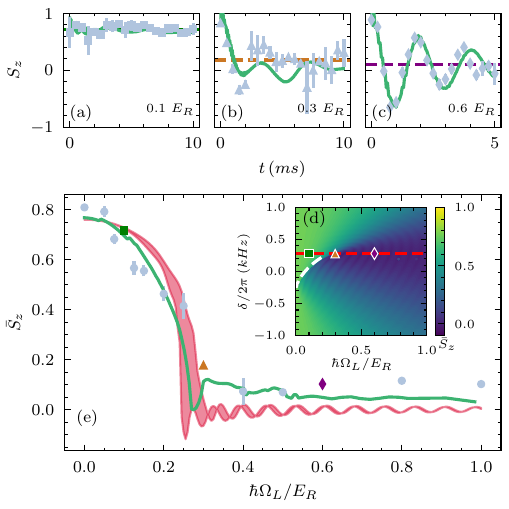}
\caption{Dynamical phase transition. (a)-(c) Individual spin oscillations over a substantial evolution time with $\hbar\Omega_{L} = 0.1$, $0.3$, and $\qty{0.6}{E_{R}}$ for (a), (b), and (c) respectively. Light blue square, triangle, and diamond dots represent experimentally collected spin-polarization averaged over three images with the error bar representing the standard deviation between them. If unseen, the standard deviation is less than the height of the point. The green curves represent the GP derived numerical time trace curves. The colored dashed lines are the total spin-polarization average of all experimental points for separate values of $\hbar\Omega_{L}$. Inset (d) Analytical phase diagram of order parameter $\bar{S}_z$ averaged over $\qty{10}{ms}$ for different values of $\hbar\Omega_{L}$ and final quench Raman detuning $\delta$. Experimental data was taken along the red dashed line which represents a cut of the phase diagram at $\delta=2\pi\times\qty{280}{Hz}$. The dashed white line indicates the theoretical phase boundary in the homogeneous limit obtained from the effective potential model~\cite{sup_mat}. (e) Cut of $\bar{S}_{z}$ along red dashed line in (d). Each point represents an experimental time average of the spin-polarization for different $\hbar\Omega_{L}$. The time domain over which we average is such that it is sufficient to resolve several oscillations between spin states~\cite{sup_mat}. The green curve represents GP numerics while the red shaded region denotes the analytic solution to the system bounded by a detuning range of $\qty{260}{Hz} \leq \delta/(2\pi) \leq \qty{300}{Hz}$. The green square, orange triangle, and purple diamond correspond to experimentally determined spin-polarization time averages of (a), (b), and (c) respectively with their error bars representing standard error between each of the three run's time-averaged spin-polarization.}
\label{fig:dpt}
\end{figure}

\paragraph{\textcolor{blue}{Dynamical phase transition.}}
Closely related to self-trapping is a pronounced change in long-term dynamics of the system, compared to one that does not exhibit self-trapping. Generally, self-trapping manifests as a slowing down or temporary plateau in the evolution of an observable within a nonlinear system. In the SOC + ML system, self-trapping can be understood as a tendency for atoms to remain in a particular spin state, with certain spin and momentum dynamics becoming damped over long timescales. The abrupt change in the long time behavior when transitioning between the self-trapping and non-self-trapping regimes signifies a dynamical phase transition.

We can induce and probe such dynamics by performing quenches of the Raman detuning. By varying the optical lattice strength, we can place the system in either the self-trapping or non-self-trapping regime. Observing the long-timescale oscillations of the spin polarization after the quench in the different regimes allows us to measure the dynamical phase transition.
We begin with a highly detuned SOC + ML BEC, initialized with $\delta = 2\pi\times\qty{5}{kHz}$ and set $\hbar\Omega_L$ to a fixed value where $\qty{0.0}{E_R} \leq \hbar\Omega_L \leq \qty{1.0}{E_R}$. We then perform a Raman frequency quench from $2\pi\times\qty{5}{kHz} \rightarrow 2\pi\times\qty{0}{Hz}$, letting the system evolve after the quench for varying amounts of time, and then follow our standard imaging procedure. Fig.~\ref{fig:dpt}(a-c) show the spin-polarization oscillations induced by the Raman detuning quench for three values of optical lattice strength, $\hbar\Omega_L = \qty{0.1}{E_R}$, $\qty{0.3}{E_R}$, and $\qty{0.6}{E_R}$. 

In our analysis, we take the time-averaged spin polarization, $\bar{S}_z = \int_0^{T}S_z \text{d}t/T$, as the order parameter of the dynamical phase transition. Averaging over a sufficiently long temporal window $T\gg\tau$, where $\tau$ is the spin oscillation period~\cite{sup_mat}, we observe a clear signature of the dynamical phase transition, as shown in Fig.~\ref{fig:dpt}(e). At high $\hbar\Omega_L$, the system shows pronounced oscillation in spin-polarization. As $\hbar\Omega_L$ decreases, the oscillations become increasingly damped, with a higher time-averaged spin polarization $\bar{S}_z$.

Numerical simulations of the homogeneous two-mode model are performed by evolving the spin populations and computing their imbalance. The resulting dynamical phase diagram from analytical solutions to the system [Fig.~\ref{fig:dpt}(d)], obtained from time-averages over $T=\qty{10}{ms}$ to match experimental conditions, agrees with the experimental observation of two regimes: a self-trapped phase and a delocalized phase. In the low-lattice, low-detuning regime near the critical lattice strength $\Omega^c_L$, the oscillation period diverges as $\tau \sim \ln|\Omega_L-\Omega_L^c|$~\cite{muniz2020, guan2021, sup_mat}, signaling a dynamical phase transition.

\begin{figure}[t]
\centering
\includegraphics[scale=1.0]{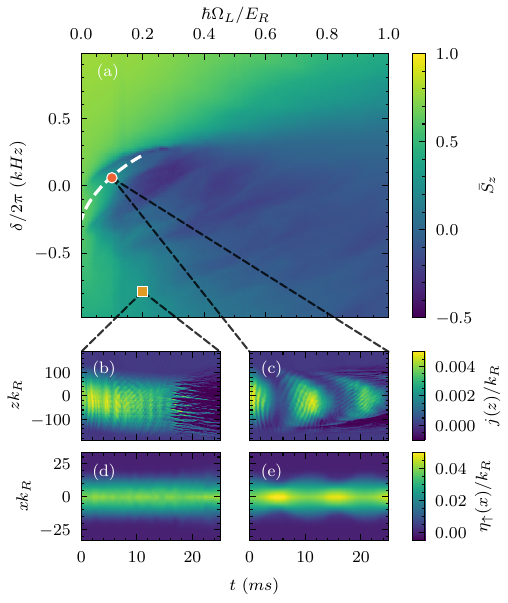}
\caption{Dynamical phase transition beyond the two-state model. (a) Dynamical phase diagram from GP simulations showing time-averaged polarization for $25$ ms runs, with the two-mode separatrix (dashed line) overlaid. Unlike the 10 ms case, the delocalized phase exhibits uniformly negative polarization. (b)-(c) Polarization density $j(z) = \int(|\psi_\uparrow|^2-|\psi_\downarrow|^2)\text{d}x\text{d}y/N$ in the delocalized phase (b) and near the boundary (c). Spin domains form after $\sim\qty{15}{ms}$, signaling loss of integrability. (d)-(e) Normalized transverse density for the spin-up component $\eta_\uparrow(x)=\int |\psi_\uparrow|^2 \text{d}y\text{d}z/N_\uparrow$ at the same points. No transverse motion is seen in the delocalized phase, while near the boundary a resonance with the transverse mode induces oscillations, consistent with the new phase boundary in (a).}
\label{fig:DPT_outlook}
\end{figure}

To test the robustness of these features, we perform GP simulations including the harmonic trap and extended the averaging window to $\qty{25}{ms}$ [Fig.~\ref{fig:DPT_outlook}(a)]. Finite-size effects modify the phase boundary and the value of $\bar{S}_z$ in the delocalized region, while the self-trapped phase remains in excellent agreement with the two-mode model. In the delocalized phase, axial defects emerge after $\sim\qty{15}{ms}$ [Fig.~\ref{fig:DPT_outlook}(b-d)], producing a dynamical instability that damps spin oscillations and invalidates the two-mode description. Notably, the phase boundary develops a cusp-shaped region [red circle in Fig.~\ref{fig:DPT_outlook}(a)], where resonance features arise and induce coupled axial and transverse oscillations [Fig.~\ref{fig:DPT_outlook}(c-e)]. A Bogoliubov analysis will identify the unstable modes and enable the construction of an extended few-mode model that captures the richer dynamical phase diagram.

\paragraph{\textcolor{blue}{Conclusion \& discussion.}}
In this work, we demonstrate self-trapping and a DPT in momentum space.  While these findings are noteworthy in their own right as they represent the first observation of MQST and DPT in a momentum space BEC, they also highlight the potential of this system as a versatile platform for future applications. For example, highly controllable DPTs based on MQST could serve as a testbed for further research in areas such as non-equilibrium criticality~\cite{zunkovic2018} and quantum enhanced metrology~\cite{macieszczak2016, guan2021}.

\begin{acknowledgments}
\textit{Acknowledgments:} C.S., C.H., A.M., and P.E. acknowledge funding from NSF through Grant No. PHY-2207588. P.E. also acknowledges funding from a Boeing Endowed Professorship. F.S. and Q.G acknowledge funding from NSF through Grants No. PHY-2409600, the WSU Claire May \& William Band Distinguished Professorship Award, and through a WSU New Faculty Seed Grant. We acknowledge insightful discussions with D. Blume.
\end{acknowledgments}

\bibliography{bibliography}

\renewcommand{\theequation}{S\arabic{equation}}
\setcounter{equation}{0}

\renewcommand{\thefigure}{S\arabic{figure}}
\setcounter{figure}{0}

\renewcommand{\thetable}{S\arabic{table}}
\setcounter{table}{0}

\section{Supplementary Materials}

\paragraph{\textcolor{blue}{Macroscopic quantum self-trapping.}}
For $\hbar\Omega_R < 4E_R$ and small $\delta$, the system can be described by a wavefunction composed of two BEC modes localized at the band minima (neglecting contributions from the excited band) and sharing the same spatial profile,
\begin{eqnarray}
    \Psi(\mathbf{r}, t) = \left[a_l(t)\chi_l e^{-i k_L z} + a_r(t)\chi_r e^{i k_L z}\right]e^{i k_b z}\varphi(\mathbf{r}),\ 
\end{eqnarray}
where $a_{l/r}(t)$ are the complex amplitudes of each mode, $\varphi(\mathbf{r})$ is the background wavefunction, and $\hbar k_b$ is a detuning-induced momentum bias. Writing $a_{l/r} = \sqrt{N_{l/r}} e^{i \theta_{l/r}}$ and defining the relative population imbalance $s_z = (N_r - N_l)/(N_r + N_l)$ and the relative phase $\theta = \theta_r - \theta_l$, the system reduces to an effective two-mode Josephson junction Hamiltonian~\cite{mukhopadhyay2024},
\begin{eqnarray}
    H_\text{eff} = -\frac{1}{2}gn\chi^2s_z^2 - \hbar\Omega_L\chi\sqrt{1-s_z^2}\cos\theta - \hbar\delta s_z,
\end{eqnarray}
where $\chi=|\chi_l^{\dagger}\chi_r|$ is proportional to the inter-mode hopping amplitude between the two band minima and $gn=gN\int|\varphi({\bf{r}})|^4\text{d}{\bf{r}}$ is the mean-field energy.

In the experiment, we measure the spin polarization $S_z=(N_\uparrow - N_\downarrow) / N = (|a_l|^2\chi_l^\dagger \sigma_z \chi_l + |a_r|^2\chi_r^\dagger\sigma_z\chi_r)/N$. At zero detuning, taking $\chi_r = -\sigma_x\chi_l$ gives
\begin{eqnarray}
    S_z = (\chi_l^\dagger \sigma_z\chi_l)s_z.
    \label{order_parameters_def}
\end{eqnarray}
For the ground state, $\chi_l = [\cos{\alpha},\ e^{i\nu}\sin{\alpha}]^T$ with $\cos{(2\alpha)}=k_L/k_R$ and with arbitrary phase $\nu$, yielding the result quoted in the main text: $S_z = (k_L/k_R)s_z$.

The equations of motion are derived via
\begin{align}
&\dot{s_z} = \frac{\partial H_{\text{eff}}}{\hbar\partial \theta} = \Omega_L\chi\sqrt{1-s_z^2}\sin\theta, \\
&\dot{\theta}  = -\frac{\partial H_{\text{eff}}}{\hbar\partial s_z} = -\frac{gn\chi^2}{\hbar} s_z + \Omega_L\chi\frac{s_z\cos\theta}{\sqrt{1-s_z^2}} - \delta,
\end{align}
where the equilibrium solutions are found using $\dot{s_z}=\dot{\theta}=0$. From these conditions, the equilibrium spin polarization $s_{z,0}$ satisfies
\begin{eqnarray}
    \frac{s_{z,0}^2}{1-s_{z,0}^2} = \frac{(gn\chi^2s_{z,0}+\hbar\delta)^2}{(\hbar\Omega_L\chi)^2},
    \label{eq:eq_condition}
\end{eqnarray}
which supports a finite $s_{z,0}$ for $\hbar\Omega_L < gn\chi$, signaling the transition between the self-trapped regime and the delocalized regime.

\begin{figure}[t]
\centering
\includegraphics[scale=1.0]{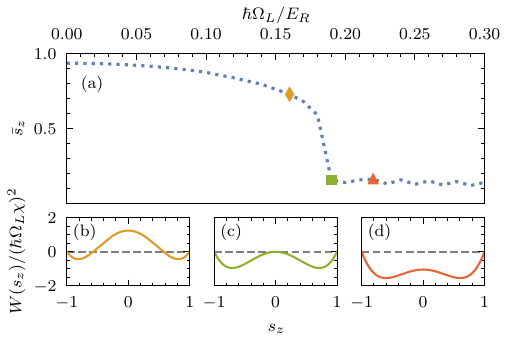}
\caption{Effective potential across the dynamical phase transition. (a) The dotted line is the time-averaged polarization $\bar{s}_z$ obtained from the model simulation at $\delta = 2\pi\times\qty{0}{Hz}$ as a function of $\hbar\Omega_L$ for a time interval $T=\qty{40}{ms}$. A dynamical phase transition is observed, with the critical point located around $\hbar\Omega_L \sim \qty{0.20}{E_R}$. The critical point in the experiments and GP simulations is higher ($\sim0.3E_R$) due to the detuning shift from the $|1,\ 1\rangle$ state. The diamond, square, and triangle correspond to the the particular values of $\hbar\Omega_L$ shown in panels (b–d).
(b–d) Effective potential $W(s_z)$ for (b) $\hbar\Omega_L = \qty{0.16}{E_R}$, (c) $\hbar\Omega_L = \qty{0.19}{E_R}$, and (d) $\hbar\Omega_L = \qty{0.22}{E_R}$.
}
\label{fig:dpt_eff_potential_model}
\end{figure}

\paragraph{\textcolor{blue}{Dynamical phase transition.}}
To probe the dynamical properties of the system, we prepare the condensate in an out-of-equilibrium initial state with population imbalance $s_z=1$. In this case, the equation of motion can be written as $(\hbar\dot{s_z})^2 + W(s_z) = 0$ where the effective trap reads~\cite{muniz2020,guan2021}
\begin{align}
\label{eq:effective_potential}
&W(s_z)=(1-s_z)\times \\\nonumber
& \Big\{(1-s_z)\Big[\tfrac{1}{2}gn\chi^2(1+s_z)+\hbar\delta\Big]^2 - (\hbar\Omega_L\chi)^2(1+s_z)\Big\}.
\end{align}
Depending on the parameters, the dynamics correspond to oscillations of $s_z$ around either a local minimum (self-trapped phase) or the global minimum (delocalized phase) of $W(s_z)$. The spin oscillation period $\tau$ between the polarization states $s_{z}'$ and $s_z''$ is obtained via
\begin{equation}
\tau = 2\hbar\int_{s_{z}'}^{s_{z,}''} \frac{s_zds_z}{\sqrt{-W(s_z)}},
\end{equation}
where the $s_{z}'$ and $s_{z}''$ are determined by the roots of $W(s_z)=0$. In general, $W(s_z)$ is a fourth-order polynomial in terms of $s_z$ which can be factorized as
\begin{equation}
W(s_z) = \frac{g^2n^2\chi^4}{4}(s_z-1)(s_z-s_z^*)(s_z-s_z^*+\varepsilon)(s_z-a),
\end{equation}
with $\{a,\ s_z^*-\varepsilon,\ s_z^*,\ 1\}$ being the roots ordered from lower to higher. Near the critical point, the dominant contribution to the integral is
\begin{equation}
\tau \propto \frac{s_z^*}{\sqrt{(s_z^*-a)(1-s_z^*)}}
\int_0^{1-s_z^*} \frac{ds_z}{\sqrt{s_z(s_z+\varepsilon)}},
\label{eq:period_dpt}
\end{equation}
where the separation between the two relevant roots scales for some value $\gamma$ as $\varepsilon \sim |\Omega_L - \Omega_L^c|^\gamma$, with $\Omega_L^c$ denoting the value of $\Omega_L$ at which the two inner roots merge as shown in Fig. \ref{fig:dpt_eff_potential_model}(c). From \eqref{eq:period_dpt} we find a logarithmic divergence of the oscillation period at criticality,
\begin{equation}
\tau \sim \ln|\Omega_L - \Omega_L^c|,
\end{equation}
which signals the presence of a dynamical phase transition.

\begin{figure}[t]
\centering
\includegraphics[scale=1.0]{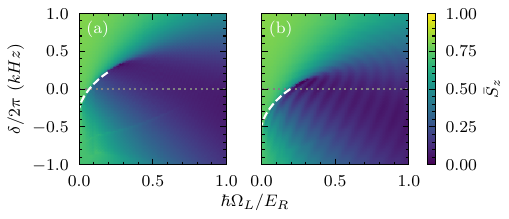}
\caption{Effects of considering a larger time window $T$ and effects from the $|1,\ 1\rangle$ state. (a) Dynamical phase diagram from the two-state model, where the time-averaged polarization is computed over a longer window ($T=\qty{40}{ms}$). The small oscillations present in the original diagram [Fig.~\ref{fig:dpt}(a)] are suppressed. (b) Dynamical phase diagram for $T=\qty{10}{ms}$ obtained without including the third state. Compared with the full three-state simulation, a shift in the detuning is observed, consistent with Eq.~\eqref{eq:detuning_shift}. In both panels, the white dashed lines indicate the theoretical phase boundary obtained from the effective potential model.}
\label{fig:dpt_role_of_third_state_time_window}
\end{figure}

In this work, the spin polarization is averaged over a time window shorter than $\qty{10}{ms}$. Two-state simulations show negligible difference between $\qty{10}{ms}$ and $\qty{40}{ms}$ windows, except for a smoothing of small oscillations [Fig.~\ref{fig:dpt_role_of_third_state_time_window}(a)], which occur because the oscillation period is comparable to the averaging window. The two-state model agrees quantitatively with GP simulations up to $\qty{10}{ms}$, beyond which additional dynamics not captured by the model become relevant.

\paragraph{\textcolor{blue}{Role of the $|1,\ 1\rangle$ hyperfine state.}}
The two-mode model discussed above involves the hyperfine states $|1,-1\rangle$ and $|1,0\rangle$. The third state of the $F=1$ hyperfine manifold,  $\ket{1,\ 1}$, is adiabatically eliminated due to the large quadratic Zeeman shift $2\hbar q_z\approx h\times 14$~kHz induced by the $10$ Gauss external magnetic field. The shift is non-negligible for certain parameter regimes considered in this work, especially in determining the critical detuning near the quantum phase transition and the dynamical phase transition.   

To account for the influence of the hyperfine state $\ket{1,\ 1}$, we extend our model to a three-component system. The extended single-particle Hamiltonian in the basis $\Psi_3 = [\psi_{\uparrow}, \psi_{\downarrow}, \phi]^T$ , where $\phi$ corresponds to the wavefunction of the state $\ket{1,\ 1}$, takes the block form,
\begin{equation}
    H_3 = 
    \begin{pmatrix}
        H_0 & C \\
        C^T & \Delta
    \end{pmatrix},
\end{equation}
where $H_0=H_{\text{SOC}} + [V_L(z) + U(\mathbf{r})]\sigma_0$ is the original $2\times2$ single-particle Hamiltonian seen in Eq.~\eqref{eq:gp_equation} of the main text, and $C = [0,\, \hbar\Omega_R/2]^T$ describes the coupling between the SOC subspace and the $\ket{1,\ 1}$ state. The $\Delta$ captures the single-particle energy of the state $\ket{1,\ 1}$ by
\begin{equation}
\label{delta}
    \Delta = \tfrac{1}{2m}\left( \hbar\mathbf{k} + 3\hbar k_R\ \mathbf{e}_z \right)^2 + \tfrac{3}{2}\hbar\delta + 2\hbar q_z + U(\mathbf{r}) + V_L(z).
\end{equation}
The time-dependent GP equation for the three components becomes,
\begin{align}
    i\hbar\frac{\partial\Psi_3}{\partial t} &= \bigg[\frac{\hbar^2}{2m}(\mathbf{k}\tilde{\sigma}_0-\tilde{\sigma}_z k_R\mathbf{e}_z)^2 
        - \tfrac{1}{2}\hbar\delta\tilde{\sigma}_z 
        + 2 \hbar q_z\tilde{\sigma}_3 + U(\mathbf{r})\tilde{\sigma}_0 \nonumber \\
    &\quad + V_L(z)\tilde{\sigma}_0 
        + \tfrac{1}{2}\hbar\Omega_R\tilde{\sigma}_x 
        + g\Psi_3^\dagger\Psi_3\bigg] \Psi_3.
\end{align}
where
\begin{align}
    \tilde{\sigma}_0 &= 
    \begin{pmatrix}
        1 & 0 & 0 \\
        0 & 1 & 0 \\
        0 & 0 & 1
    \end{pmatrix}, \\[6pt]
    \tilde{\sigma}_z &= 
    \begin{pmatrix}
        1 & 0 & 0 \\
        0 & -1 & 0 \\
        0 & 0 & -3
    \end{pmatrix}, \\[6pt]
    \tilde{\sigma}_x &= 
    \begin{pmatrix}
        0 & 1 & 0 \\
        1 & 0 & 1 \\
        0 & 1 & 0
    \end{pmatrix}, \\[6pt]
    \tilde{\sigma}_3 &= 
    \begin{pmatrix}
        0 & 0 & 0 \\
        0 & 0 & 0 \\
        0 & 0 & 1
    \end{pmatrix}.
\end{align}
The term $q_z$ represents the quadratic Zeeman shift. For $\ket{1,\ 1}$, $\hbar q_z \sim h \times \qty{7}{kHz}$, which is much larger than typical values of $\hbar\Omega_R$, $\hbar\delta$, and the lattice depth $\hbar\Omega_L$. 

To obtain an effective Hamiltonian for the low-energy subspace, we employ the Schrieffer--Wolff transformation, a perturbative technique that systematically eliminates the high-energy degrees of freedom. The time-dependent Schrödinger equation then decomposes into the coupled equations:
\begin{eqnarray}
\label{schrodinger_equation_psi}
    i\hbar \frac{\partial \Psi}{\partial t} &=& H_0\Psi + \frac{\hbar\Omega_R}{2}\begin{pmatrix}
        0 \\ \phi
    \end{pmatrix}, \\
    i\hbar \frac{\partial\phi}{\partial t} &=& \Delta\,\phi + \frac{\hbar\Omega_R}{2}\psi_\downarrow.
    \label{schrodinger_equation_phi}
\end{eqnarray}
Given that $\Delta$ is significantly larger than all other relevant energy scales, and that the population in the auxiliary state $\phi$ remains negligible over relevant timescales, we assume that the dynamics of $\phi$ are fast and adiabatically follow the low-energy subspace. Consequently, we set $\partial\phi/\partial t = 0$ in Eq.~\eqref{schrodinger_equation_phi}, and neglect the kinetic term in Eq.~\eqref{delta}. This allows us to solve algebraically for $\phi$ and substitute the result back into Eq.~\eqref{schrodinger_equation_psi}, yielding an effective two-component equation:
\begin{equation}
    i\hbar \frac{\partial \Psi_3}{\partial t} = \left[H_0 + \frac{(\hbar\Omega_R)^2}{8\Delta}\sigma_z\right]\Psi_3.
\end{equation}
The resulting effective Hamiltonian retains the original spin-orbit coupling and optical lattice structure, and includes a shift of the detuning due to virtual coupling to the third hyperfine state
\begin{eqnarray}
    \delta \longrightarrow \delta +  \frac{\hbar\Omega_R^2}{6\hbar\delta + 8\hbar q_z + 36\hbar^2k_R^2/2m}.
\label{eq:detuning_shift}
\end{eqnarray}

Since the shift scales inversely with $\Delta$, it becomes more pronounced at lower values of $\Delta$. In our system, the minimum value that $\Delta$ can take is $\Delta_{\text{min}} = 9\hbar^2k_R^2/2m + 2\hbar q_z \sim 16\, E_R$, which leads to a detuning shift of up to $2\pi\times\qty{223}{Hz}$ for a Raman coupling strength of $\hbar\Omega_R = \qty{2.7}{E_R}$. Fig.~\ref{fig:dpt_role_of_third_state_time_window}(b) shows the phase diagram obtained by the two-state simulations without including the third state. The dashed white line is the phase boundary prediction by Eq.~\eqref{eq:effective_potential}. In contrast, The phase boundary in Fig.~\ref{fig:dpt_role_of_third_state_time_window}(a) is obtained after including the shift given in Eq.\eqref{eq:detuning_shift}.

\begin{figure}[t]
\centering
\includegraphics[scale=1.0]{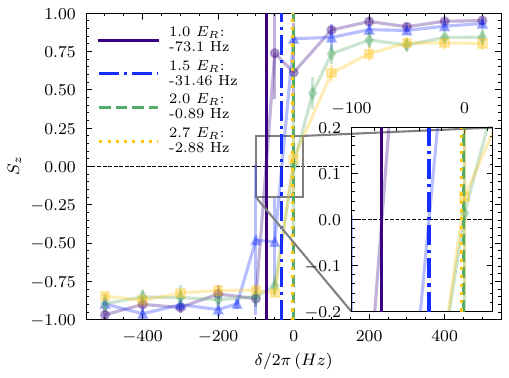}
\caption{Influence of the off-resonant $|1,1\rangle$ state on the ground state of the system. The purple circles, blue triangles, green diamonds, and yellow squares represent experimental data taken at $\hbar\Omega_R=$ 1.0, 1.5, 2.0, and $\qty{2.7}{E_R}$ respectively. The vertical lines - purple solid, blue dot-dashed, green dashed, and yellow dotted - correspond to where the interpolated experimental $S_z=0$.
}
\label{fig:3rdStGNDInfluence}
\end{figure}

In order to experimentally measure the effect of the off-resonant $|1,1\rangle$ state on the SOC ground state, we prepared the SOC system without the matching optical lattice by adiabatically ramping on the strength of the Raman beams to be 1.0, 1.5, 2.0, or $\qty{2.7}{E_{\text{R}}}$. For each value of $\hbar\Omega_{\text{R}}$, we set $\hbar\delta$ to be values that span between $-h\times 500\text{Hz}$ and $h\times500\text{Hz}$. We then hold the atoms in trap until the spin composition of the cloud reaches a steady-state after which we measure $S_{z}$. 

Fig.~\ref{fig:3rdStGNDInfluence} shows the effect of the off-resonant $|1,1\rangle$ state on the ground state of the SOC system by measuring the zero-detuning offset point of the cloud's spin polarization. The zero-detuning is calibrated (the calibration procedure is discussed in detail in the following section labeled ``Zero-detuning calibration procedure'') at $\hbar\Omega_{\text{R}} = \qty{2.7}{E_{\text{R}}}$ and no optical lattice. At lower values of $\hbar\Omega_{\text{R}}$ we see a clear shift in the value of Raman detuning at which the spin-polarization crosses zero. At higher Raman strengths, the interaction between the far-detuned $|1,1\rangle$ state and the atoms increase, causing an effective detuning which can be seen to differ from the detuning at lower $\hbar\Omega_{R}$.

\begin{figure}[h]
\centering
\includegraphics[scale=1.0]{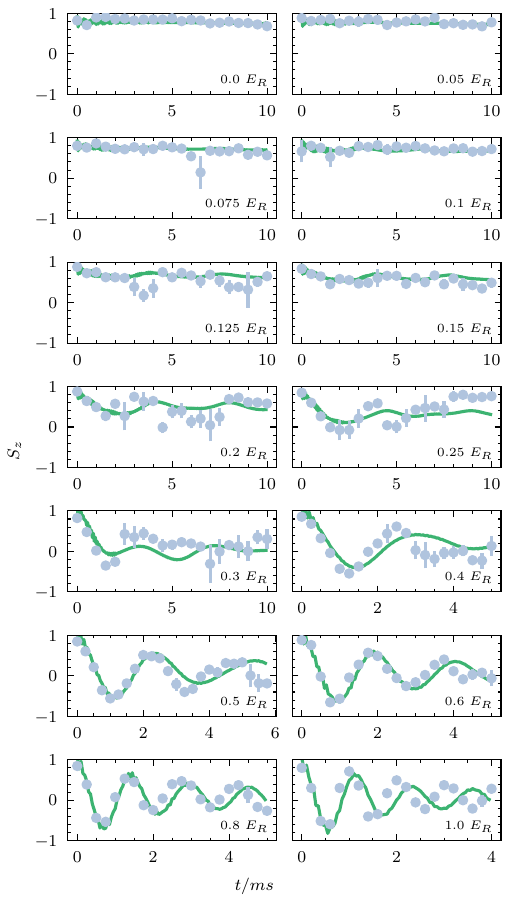}
\caption{Time dynamics of quenched SOC system. Gray dots correspond to experimental data while green curves show GP numerical time traces. Each panel consists of a quench done with fixed value of $\hbar\Omega_{L}$ where $\qty{0.0}{E_R} \leq \hbar\Omega_{L} \leq \qty{1.0}{E_{R}}$.}
\label{fig:Quench_Oscillations}
\end{figure}

\paragraph{\textcolor{blue}{Numerical details.}}
We numerically solve the three-dimensional GP equation \eqref{eq:gp_equation} under axial symmetry using a fourth-order Runge–Kutta (RK4) scheme. The simulation domain has dimensions $L_\perp \times L_z = (33.88 \times 377.37)/k_R^2$, discretized on a grid of $N_\perp \times N_z = 64 \times 512$ points. The trapping potential is characterized by $(\omega_x,\ \omega_y,\ \omega_z) = 2\pi \times (176.22,\ 198.53,\ 27.99)$ Hz, for the GP simulations we use an effective transverse frequency $\omega_\perp = 2\pi\times \sqrt{176.22\times198.53}$ Hz, and the condensate contains $N \approx 1.93\times10^5$ atoms, corresponding to a healing length $\xi \simeq 1.22/k_R$, which is larger than the longitudinal grid spacing $dz=0.74/k_R$. Time evolution is performed with a step size $\Delta t = 0.01\hbar/E_R$, and the ground state is obtained via imaginary-time propagation with long enough cooling time. For example, the ground state at $\hbar\Omega_R = 2.7E_R$ and $\hbar\delta = 2.55E_R$ yields a nonlinear interaction factor $gn = 0.60E_R$ and an overlap $|\chi_l^{\dagger}\chi_r| = 0.65$.

\paragraph{\textcolor{blue}{Choice of experimental temporal domain in quenched SOC system.}}
Important to the discussion of DPTs is over how long a temporal window we should average to obtain $\bar{S}_{z}$. As seen in~\ref{fig:dpt_role_of_third_state_time_window}, the time averaged spin dynamics are susceptible to slight change given different averaging times. This can be visualized quite easily considering the spin dynamics for $\hbar\Omega_{L}\gg\qty{0.3}{E_R}$. Given that the spin dynamics follow sinusoidal behavior for high optical lattice strengths as can be seen in Fig.~\ref{fig:Quench_Oscillations}, choosing a time span that encompasses only a portion of one oscillation would inevitably skew $\bar{S}_{z}$. In order to average the spin dynamics properly, we make sure the time domain over which we are averaging spans several full oscillations.

For high optical lattice strength, $\hbar\Omega_{L}>\qty{0.3}{E_R}$, the temporal span chosen ranges from $\qty{4}{ms}$ to $\qty{6}{ms}$. Intuitively, as $\hbar\Omega_L$ becomes larger, so does the coupling frequency between spin states. For larger $\hbar\Omega_L$ we can therefore average over less time while still being able to resolve several complete oscillations between spin states. At lower $\hbar\Omega_L$ however, where self-trapping dominates the system, we see only a partial spin oscillation or none at all before the spin mixture stagnates at a fixed ratio. For $\hbar\Omega_L \leq \qty{0.3}{E_R}$ we find that the long time spin dynamics settle into a steady-state mixture by 10 ms, allowing us to average over this time scale without losing the defining characteristics of the DPT.  

\paragraph{\textcolor{blue}{Zero-detuning calibration procedure.}}
We are able to engineer specific energy differences between the two states in the SOC system by precisely controlling the relative detuning of the Raman beams. We calibrate the relative detuning of the Raman beams to within 10 Hz by adiabatically ramping on the Raman beams within the presence of a BEC and looking for which values of Raman frequency detuning we observe a mixed spin state in our BEC. This mixed state condition only occurs for a very narrow range of Raman frequency detuning where the two momentum states minima are at equal energy and $\delta = 2\pi\times \qty{0}{Hz}$. Once we calibrate the relative detuning of the Raman beams, we relate all of the data taken immediately afterwards to this calibrated frequency detuning value where any offset from the calibrated value corresponds to an offset in $\delta$. Since there is a natural long time scale drift in the energy difference between the two spin states due to magnetic and thermal noise, we perform a zero-detuning calibration after every nine experimental runs of the SOC system. Experiments utilizing a SOC + ML BEC have a Raman detuning uncertainty of $\delta = \pm 2\pi\times\qty{100}{Hz}$.

\end{document}